\documentclass[11pt,a4paper]{article}
\usepackage[english]{babel}
\usepackage{amsmath,amsthm,amssymb,epsfig,latexsym}
\usepackage{graphics}
\newcommand{\so}{\scriptscriptstyle \rm I}
\newcommand{\st}{\scriptscriptstyle \rm I\hspace{-1pt}I}

\newcommand{\be}[1]{\begin{equation}\label{#1}}
\newcommand{\ba}[1]{\begin{multline}\label{#1}}
\newcommand{\ee}{\end{equation}}
\newcommand{\ea}{\end{eqnarray}}

\newcommand{\num}{\\\rule{0pt}{20pt}}

\newcommand{\dis}{\displaystyle}

\newcommand{\Res}{\mathop{\rm Res}}

\newtheorem{prop}{Proposition}[section]

 \makeatletter
 \@addtoreset{equation}{section}
 \makeatother

\begin{document}

\begin{flushright}
LAPTH-028/12
\end{flushright}

\vspace{20pt}

\begin{center}
\begin{LARGE}
{\bf Highest coefficient of scalar products in $SU(3)$-invariant integrable models}
\end{LARGE}

\vspace{40pt}

\begin{large}
{S.~Belliard${}^a$, S.~Pakuliak${}^b$, E.~Ragoucy${}^c$, N.~A.~Slavnov${}^d$\footnote[1]{samuel.belliard@univ-montp2.fr, pakuliak@theor.jinr.ru, eric.ragoucy@lapp.in2p3.fr, nslavnov@mi.ras.ru}}
\end{large}

 \vspace{12mm}

${}^a$ {\it  Universit\'e Montpellier 2, Laboratoire Charles Coulomb,\\ UMR 5221,
F-34095 Montpellier, France}

\vspace{4mm}

${}^b$ {\it Laboratory of Theoretical Physics, JINR, 141980 Dubna, Moscow reg., Russia,\\
Moscow Institute of Physics and Technology, 141700, Dolgoprudny, Moscow reg., Russia,\\
Institute of Theoretical and Experimental Physics, 117259 Moscow, Russia}

\vspace{4mm}

${}^c$ {\it Laboratoire de Physique Th\'eorique LAPTH, CNRS and Universit\'e de Savoie,\\
BP 110, 74941 Annecy-le-Vieux Cedex, France}

\vspace{4mm}

${}^d$ {\it Steklov Mathematical Institute,
Moscow, Russia}

\end{center}

\vspace{2mm}

\begin{abstract}
We study $SU(3)$-invariant integrable models solvable by nested algebraic Bethe ansatz. Scalar
products of Bethe vectors in such models can be expressed in terms of a bilinear combination of their
highest coefficients. We obtain various different representations for the highest coefficient in terms
of sums over partitions. We also obtain multiple integral representations for the highest coefficient.
\end{abstract}

\vspace{2mm}

\section{Introduction}

The problem of calculating  local operators form factors
and correlation functions in quantum integrable models is of highest importance. When
integrable models are solvable by  algebraic Bethe ansatz \cite{FadST79,BogIK93L,FadLH96} this
problem can be reduced to the calculation of scalar products  of Bethe vectors.

The scalar products of Bethe vectors were first considered for  $\frak{gl}_2$-based integrable models \cite{Kor82,IzeK84}.
In these works the notion of  highest coefficient $K_n$
of a scalar product was introduced. Any scalar
product can be expressed in terms of a bilinear combination of  $K_n$ (Izergin--Korepin formula).
It was shown in \cite{Kor82,IzeK84} that for the models with $SU(2)$-symmetry (and $q$-deformed $SU(2)$-symmetry)
$K_n$ is equal to the partition function of the six-vertex model with domain wall boundary conditions.
An explicit determinant representation for this partition function was derived in \cite{Ize87}.

A wide class of quantum integrable models is associated with higher rank algebras $\frak{gl}_N$.
An algebraic Bethe ansatz for these type of models is called hierarchical (or nested) and was
introduced in \cite{KulR83} (see also \cite{BelR08}). The first
results concerning the scalar products in the models with $SU(3)$-invariant
$R$-matrix was obtained by N.Yu.~Reshetikhin in \cite{Res86}. There, an analog of Izergin--Korepin
formula  for the scalar product of generic Bethe vectors  and a determinant representation
for the norm of the transfer-matrix eigenvectors were found. Similarly to the Izergin--Korepin
formula Reshetikhin's representation for the scalar product  can be considered as a bilinear
combination of highest coefficients ($Z_{a,b}$). In turn, $Z_{a,b}$ is equal to a special
partition function. The study of this partition function is the subject of the present paper.

Recently the explicit representation for the $Z_{a,b}$ associated with $SU(3)$-invariant
$R$-matrix was obtained in  \cite{Whe12}. There, $Z_{a,b}$ was given as a trilinear
combination of $K_n$. There exist, however, many other representations of similar type. We have
found it very useful to use different representations for the $Z_{a,b}$ in studying the
problem of scalar products. In particular, this approach allowed us to derive a determinant
representation for the scalar product of eigenvectors of the transfer-matrix  and twisted
transfer-matrix (see our forthcoming publication \cite{BelPRS12}). In the present paper we
prove several representations for $Z_{a,b}$ in terms of sums over partitions and in terms of
multiple integrals of Cauchy type.

The article is organized as follows. In Section~\ref{S-DefNot} we give the definition of
the partition function equivalent to $Z_{a,b}$ and explain the notations used below. Section~\ref{S-MRes} gathers our results:
first, we give a list of sum formulas for the highest coefficient $Z_{a,b}$ (section \ref{S-Rep-Part}), then we provide integral representations for $Z_{a,b}$ (section \ref{S-Int-Part}), and finally, we show recursion relations on the highest coefficient $Z_{a,b}$, that allow one to fix it
unambiguously (section \ref{S-RecHC-3}). The following sections deal with the proofs of our results:
in Sections~\ref{S-proof1} and \ref{S-proof2} we prove the different representations given in
Section~\ref{S-MRes}, and in Section~\ref{S-prf-RecHC-3} we prove the recursion relations. Some properties of the highest coefficient $K_n$, needed for our calculations, are given in Appendix~\ref{A-IHC}. In Appendix~\ref{A-UP}
we prove the absence of contribution of certain poles in the integral representations for $Z_{a,b}$.

\section{Definitions and notations\label{S-DefNot}}

The  $SU(3)$-invariant $R$-matrix has the form
 \be{R-mat}
 R(x,y)=\mathbf{I}+g(x,y)\mathbf{P}, \qquad g(x,y)=\frac{c}{x-y},
 \ee
where $\mathbf{I}$ is the identity matrix, $\mathbf{P}$ is the permutation matrix, $c$ is a constant.
Keeping in mind possible generalization of our results to the models with $q$-deformed $SU(3)$-symmetry
we do not stress that the function $g(x,y)$ depends on the difference $x-y$.

Apart from the function $g(x,y)$ we also introduce a function $f(x,y)$ as
\be{univ-not}
 f(x,y)=\frac{x-y+c}{x-y}.
\ee
Clearly in our case $f(x,y)=1+g(x,y)$, however it is no more true in the $q$-deformed case. Two other
auxiliary functions will be also used
\be{desand}
h(x,y)=\frac{f(x,y)}{g(x,y)}=\frac{x-y+c}{c},\qquad  t(x,y)=\frac{g(x,y)}{h(x,y)}=\frac{c^2}{(x-y)(x-y+c)}.
\ee
The following obvious properties of the functions introduced above are useful
 \be{propert}
 g(x,y)=-g(y,x),\quad h(x-c,y)=g^{-1}(x,y),\quad  f(x-c,y)=f^{-1}(y,x),\quad  t(x-c,y)=t(y,x).
 \ee

The $R$-matrix \eqref{R-mat} satisfies Yang--Baxter equation
\be{Y-B}
R_{12}(x,y)R_{13}(x,z)R_{23}(y,z)=R_{23}(y,z)R_{13}(x,z)R_{12}(x,y).
\ee
The equation \eqref{Y-B} holds in the tensor product $\mathbb{C}^3\otimes\mathbb{C}^3\otimes\mathbb{C}^3$.
The subscripts of the $R$-matrices in \eqref{Y-B} show the spaces where the given $R$-matrix acts
non-trivially.

In order to define the partition function, which is equivalent to $Z_{a,b}$ we use graphical
representation of the $R$-matrix (see \cite{Res86} for details). We picture the $R(x,y)$ by a vertex, in which the horizontal and vertical lines are associated with the spectral parameters $x$ and $y$ respectively (see Fig.~\ref{F-R-matrix}).
The edges of the vertex are labeled by the matrix indices of $(R)_{jk,\ell m}$.

We also consider the $R$-matrix $R^{t_1}(y,x)$, where $t_1$ means the transposition with respect
to the first space. This $R$-matrix is denoted by a dotted vertex (see Fig.~\ref{F-R-matrix}).

\begin{figure}[h!]
\begin{picture}(400,70)
\put(350,0){%
\begin{picture}(70,70)
\put(30,10){\line(0,1){40}}
\put(10,30){\line(1,0){40}}
\put(26,26){$\bullet$}
\put(33,10){$\scriptstyle \ell$}
\put(33,45){$\scriptstyle m$}
\put(10,33){$\scriptstyle j$}
\put(48,33){$\scriptstyle k$}
\put(28,60){$y$}
\put(-5,28){$x$}
\put(70,26){$= \bigl(R^{t_1}(y,x)\bigr)_{jk,\ell m}$}
\end{picture}
}
\put(100,0){%
\begin{picture}(70,70)
\put(30,10){\line(0,1){40}}
\put(10,30){\line(1,0){40}}
%
\put(33,10){$\scriptstyle \ell$}
\put(33,45){$\scriptstyle m$}
\put(10,33){$\scriptstyle j$}
\put(48,33){$\scriptstyle k$}
\put(28,60){$y$}
\put(-5,28){$x$}
\put(70,26){$= \bigl(R(x,y)\bigr)_{jk,\ell m}$}
\end{picture}
}
\end{picture}
\caption{\label{F-R-matrix} Graphical pictures of $R(x,y)$ and $R^{t_1}(y,x)$.}
\end{figure}

Due to \eqref{R-mat} there exists three types of vertices corresponding to non-zero entries
of $R(x,y)$ or $R^{t_1}(y,x)$. Following Baxter's terminology we call these vertices
$a$-type, $b$-type, and $c$-type \cite{Bax82}. The $a$-type vertex has all four indices equal to each other:
$j=k=\ell=m$. The corresponding statistical weights are equal to $f(x,y)$ for usual vertex and
$f(y,x)$ for dotted vertex (see Fig.~\ref{F-abc-types}, Fig.~\ref{F-abc-types-d}). For the $b$-type vertex, we have $j=k$, $\ell=m$, $j\ne\ell$ and statistical weights are equal to $1$.
Finally $j=\ell$, $k=m$, $j\ne k$  for the $c$-type vertex and
$j=m$, $k=\ell$, $j\ne k$  for the $c$-type dotted vertex. The statistical weights are
$g(x,y)$ and $g(y,x)$ respectively.

\begin{figure}[h!]
\begin{picture}(500,70)
\put(50,0){%
\begin{picture}(70,70)
\put(30,10){\line(0,1){40}}
\put(10,30){\line(1,0){40}}
%
\put(33,10){$\scriptstyle j$}
\put(33,45){$\scriptstyle j$}
\put(10,33){$\scriptstyle j$}
\put(48,33){$\scriptstyle j$}
\put(28,60){$y$}
\put(-5,28){$x$}
\put(60,28){$=f(x,y)$}
\put(0,-10){$a$-type vertex}
\end{picture}
}
\put(270,0){%
\begin{picture}(70,70)
\put(30,10){\line(0,1){40}}
\put(10,30){\line(1,0){40}}
%
\put(33,10){$\scriptstyle j$}
\put(33,45){$\scriptstyle j$}
\put(10,33){$\scriptstyle k$}
\put(48,33){$\scriptstyle k$}
\put(28,60){$y$}
\put(-5,28){$x$}
\put(60,28){$=1$}
\put(0,-10){$b$-type vertex}
\end{picture}
}
\put(460,0){%
\begin{picture}(70,70)
\put(30,10){\line(0,1){40}}
\put(10,30){\line(1,0){40}}
%
\put(33,10){$\scriptstyle j$}
\put(33,45){$\scriptstyle k$}
\put(10,33){$\scriptstyle j$}
\put(48,33){$\scriptstyle k$}
\put(28,60){$y$}
\put(-5,28){$x$}
\put(60,28){$=g(x,y)$}
\put(0,-10){$c$-type vertex}
\end{picture}
}
\end{picture}
\caption{\label{F-abc-types} $a$, $b$, $c$  vertices and their statistical weights.}
\end{figure}

\begin{figure}[h!]
\begin{picture}(500,70)
\put(50,0){%
\begin{picture}(70,70)
\put(30,10){\line(0,1){40}}
\put(10,30){\line(1,0){40}}
\put(26,26){$\bullet$}
\put(33,10){$\scriptstyle j$}
\put(33,45){$\scriptstyle j$}
\put(10,33){$\scriptstyle j$}
\put(48,33){$\scriptstyle j$}
\put(28,60){$y$}
\put(-5,28){$x$}
\put(60,27){$=f(y,x)$}
\put(0,-10){$a$-type vertex}
\end{picture}
}
\put(270,0){%
\begin{picture}(70,70)
\put(30,10){\line(0,1){40}}
\put(10,30){\line(1,0){40}}
\put(26,26){$\bullet$}
\put(33,10){$\scriptstyle j$}
\put(33,45){$\scriptstyle j$}
\put(10,33){$\scriptstyle k$}
\put(48,33){$\scriptstyle k$}
\put(28,60){$y$}
\put(-5,28){$x$}
\put(60,27){$=1$}
\put(0,-10){$b$-type vertex}
\end{picture}
}
\put(460,0){%
\begin{picture}(70,70)
\put(30,10){\line(0,1){40}}
\put(10,30){\line(1,0){40}}
\put(26,26){$\bullet$}
\put(33,10){$\scriptstyle k$}
\put(33,45){$\scriptstyle j$}
\put(10,33){$\scriptstyle j$}
\put(48,33){$\scriptstyle k$}
\put(28,60){$y$}
\put(-5,28){$x$}
\put(60,27){$=g(y,x)$}
\put(0,-10){$c$-type vertex}
\end{picture}
}
\end{picture}
\caption{\label{F-abc-types-d} $a$, $b$, $c$  dotted vertices and their statistical weights.}
\end{figure}

Before giving the definition of the highest coefficient $Z_{a,b}$, we describe the notations used
below. We always denote sets of variables by bar: $\bar x$, $\bar y$, $\bar w$ etc.
Individual elements of the sets are denoted by subscripts: $x_k$, $w_j$ etc. As a rule, the number of elements in the
sets is not shown explicitly in the notations, however we give a special
comments on it.  Subsets of variables are denoted by roman subscripts: $\bar x_{\so}$, $\bar t_{\rm ii}$
etc.  A special notations $\bar t_{\check k}$, $\bar x_{\check p}$ etc. are used for the sets
$\bar t\setminus t_k$, $\bar x\setminus x_p$ etc.

The partition function introduced by Reshetikhin  depends on four sets of variables. We denote it by
$Z_{a,b}(\bar t;\bar x|\bar s;\bar y)$. The subscripts show that $\#\bar t=\#\bar x=a$ and
$\#\bar s=\#\bar y=b$. We separate the sets with the same number of elements by semicolon in order to stress that $Z_{a,b}$ is not symmetric with respect to the changing of their
order, for instance, $Z_{a,b}(\bar t;\bar x|\bar s;\bar y)\ne Z_{a,b}(\bar x;\bar t|\bar s;\bar y)$.
The graphical representation of the function $Z_{a,b}(\bar t;\bar x|\bar s;\bar y)$ is
shown on Fig.~\ref{F-ParFun-def}.


\begin{figure}[h!]
\noindent
\hspace{3cm}\parbox[b][34mm][t]{40mm}{$\dis Z_{a,b}(\bar t;\bar x|\bar s;\bar y)\quad=$}
\parbox[b][6cm][t]{40mm}{\begin{picture}(200,200)
\put(40,-10){%
\begin{picture}(200,200)
\multiput(20,0)(30,0){6}{\line(0,1){170}}
\multiput(10,10)(0,30){6}{\line(1,0){170}}
\multiput(16,6)(0,30){6}{$\bullet$}
\multiput(46,6)(0,30){6}{$\bullet$}
\multiput(76,6)(0,30){6}{$\bullet$}
\multiput(23,-5)(30,0){3}{$\scriptstyle 3$}
\multiput(113,-5)(30,0){3}{$\scriptstyle 2$}
\multiput(4,8)(0,30){6}{$\scriptstyle 2$}
\multiput(183,8)(0,30){3}{$\scriptstyle 3$}
\multiput(183,98)(0,30){3}{$\scriptstyle 1$}
\multiput(23,172)(30,0){3}{$\scriptstyle 2$}
\multiput(113,172)(30,0){3}{$\scriptstyle 1$}
\put(110,180){$\overbrace{\hphantom{\hspace{1.5cm}}}^{\dis\bar t}$}
\put(20,180){$\overbrace{\hphantom{\hspace{1.5cm}}}^{\dis\bar y}$}
\put(187,35){$\left.\rule{0mm}{9mm}\right\}{\bar s}$}
\put(187,125){$\left.\rule{0mm}{9mm}\right\}{\bar x}$}
\end{picture}
}
\end{picture}}
\caption{\label{F-ParFun-def} The partition function $Z_{a,b}(\bar t;\bar x|\bar s;\bar y)$.
Vertical lines correspond to the parameters $\bar y$ and $\bar t$, horizontal lines correspond to the parameters
$\bar s$ and $\bar x$. Thus, every vertex is labeled by a pair of variables: $(x_j, y_k)$, $(s_j,t_k)$ and so on.
The vertices corresponding to the set $\bar y$ are dotted.}
\end{figure}

As usual
\be{def-Z1}
Z_{a,b}(\bar t;\bar x|\bar s;\bar y)=\sum_{\mathrm{configurations}}\quad
\prod_{\mathrm{vertices}} L(\mathrm{vertex}),
\ee
where $L(\mathrm{vertex})$ is the statistical weight corresponding to given vertex.

To conclude this section we introduce one more convention concerning the notations.
In order to avoid too cumbersome formulas below we use shorthand notations for the products of
functions $g(x,y)$, $f(x,y)$, $h(x,y)$, and $t(x,y)$. Originally these functions depend on two variables.
We use the notations $g(\bar x,\bar y)$, $f(t_k,\bar y)$ etc. for
the products of these functions with respect to the corresponding sets. For example,
 \be{SH-prod}
 \begin{array}{l}
 {\dis h(\bar y,\bar s)=\prod_{y_j\in\bar y}\prod_{s_k\in\bar s} h(y_j,s_k);\qquad
 g(x_k, \bar w)= \prod_{w_j\in\bar w} g(x_k, w_j);}\num
 {\dis f( \bar x_{\check p},x_{p})=\prod_{x_j\in\bar x\setminus x_p} f(x_j,x_p);\qquad
 f(\bar s_{\st},\bar s_{\so})=\prod_{s_j\in\bar s_{\st}}\prod_{s_k\in\bar s_{\so}} f(s_j,s_k).}
 \end{array}
 \ee

\section{Main results\label{S-MRes}}

As we have mentioned already there exists several different representations for  $Z_{a,b}$. At this time,
it is not clear to us, which one will be the most convenient for further work. Therefore we
give a whole list of different representations: hopefully the right one will be among them.

First of all we recall the determinant formula for $K_n$
(or, what is the same, for the partition function of the six-vertex model
with domain wall boundary conditions)  \cite{Ize87}. We denote it by $K_n(\bar x|\bar y)$.
The subscript $n$ means that $\#\bar x=\#\bar y=n$.  $K_n$ is given by
\begin{equation}\label{K-def}
K_n(\bar x|\bar y)
=\Delta'_n(\bar x)\Delta_n(\bar y)h(\bar x,\bar y)
\det_n t(x_j,y_k),
\end{equation}
where
\be{def-Del}
{\Delta}_n(\bar y)=\prod_{j<k}^n g(y_j,y_k),\qquad \Delta'_n(\bar x)
=\prod_{j>k}^n g(x_j,x_k).
\ee
All representations for $Z_{a,b}$ involve $K_n$.

\subsection{Sum formulas\label{S-Rep-Part}}

We first give several formulas for $Z_{a,b}$  in terms of sums over
partitions of certain sets. In all the representations given below two sets of arguments are fixed, while
the two other sets are divided into subsets.

\begin{itemize}

\item {\it The sum over partitions of $\bar s$ and $\bar x$.}
 \be{RHC-IHC}
  Z_{a,b}(\bar t;\bar x|\bar s;\bar y)=(-1)^b\sum
 K_b(\bar s-c|\bar w_{\so})K_a(\bar w_{\st}|\bar t)
  K_b(\bar y|\bar w_{\so})f(\bar w_{\so},\bar w_{\st}).
 \ee
Here $\bar w=\{\bar s,\;\bar x\}$. The sum is taken with respect to partitions of the set $\bar w$ into
subsets $\bar w_{\so}$ and $\bar w_{\st}$ with $\#\bar w_{\so}=b$ and $\#\bar w_{\st}=a$.

There exists slightly different representation, so-called  twin formula:
 \be{RHC-IHC-twin}
 Z_{a,b}(\bar t;\bar x|\bar s;\bar y)=(-1)^a\sum
  K_a(\bar w_{\st}-c|\bar x)K_a(\bar w_{\st}|\bar t)
  K_b(\bar y|\bar w_{\so})f(\bar w_{\so},\bar w_{\st}).
    \ee
All the notations are the same as in \eqref{RHC-IHC}.  If we set explicitly $w_{\so}=\{\bar s_{\so},\;\bar x_{\st}\}$
and $\bar w_{\st}=\{\bar s_{\st},\;\bar x_{\so}\}$ with $\#\bar s_{\st}=\#\bar x_{\st}=k$, then the equivalence
of \eqref{RHC-IHC} and \eqref{RHC-IHC-twin} becomes evident. Indeed, we have due to
\eqref{K-K}
 \begin{multline}\label{Rav}
 (-1)^a K_a(\bar w_{\st}-c|\bar x)= (-1)^a K_a(\bar s_{\st}-c, \bar x_{\so}-c|\bar x)=
 (-1)^k  K_k(\bar s_{\st}-c|\bar x_{\st})\\
 =(-1)^b K_b(\bar s-c|\bar s_{\so}, \bar x_{\st})=(-1)^b K_b(\bar s-c|\bar w_{\so}).
 \end{multline}
The representation \eqref{RHC-IHC} with specification $w_{\so}=\{\bar s_{\so},\;\bar x_{\st}\}$
and $\bar w_{\st}=\{\bar s_{\st},\;\bar x_{\so}\}$ was proved in \cite{Whe12}.

\item {\it The sum over partitions of $\bar y$ and $\bar t$.}
 \be{Al-RHC-IHC}
 Z_{a,b}(\bar t;\bar x|\bar s;\bar y)=(-1)^af(\bar y,\bar x)f(\bar s,\bar t)
   \sum K_a(\bar t-c|\bar\eta_{\so})K_a(\bar x|\bar\eta_{\so})K_b(\bar\eta_{\st}-c|\bar s)f(\bar\eta_{\so},\bar\eta_{\st}).
      \ee
Here $\bar\eta=\{\bar y+c,\;\bar t\}$. The sum is taken with respect to partitions of the set $\bar\eta$ into
subsets $\bar\eta_{\so}$ and $\bar\eta_{\st}$ with $\#\bar\eta_{\so}=a$ and $\#\bar\eta_{\st}=b$. This formula
also has a twin
 \be{Al-RHC-IHC-twin}
 Z_{a,b}(\bar t;\bar x|\bar s;\bar y)=(-1)^bf(\bar y,\bar x)f(\bar s,\bar t)
   \sum
 K_b(\bar\eta_{\st}-c|\bar y+c)K_a(\bar x|\bar\eta_{\so})K_b(\bar\eta_{\st}-c|\bar s)f(\bar\eta_{\so},\bar\eta_{\st}).
     \ee

\item {\it The sum over partitions of $\bar t$ and $\bar x$.}
 \ba{GF}
 Z_{a,b}(\bar t;\bar x|\bar s;\bar y)=\sum (-1)^{n}f(\bar s,\bar t_{\so}) f(\bar y,\bar x_{\st})
 f(\bar t_{\so},\bar t_{\st})f(\bar x_{\st},\bar x_{\so})\\
 \times K_n(\bar x_{\so}|\bar t_{\so})K_{a-n}(\bar x_{\st}|\bar t_{\st}-c)
 K_{b+n}(\bar y,\bar t_{\so}-c|\bar s,\bar x_{\so}).
 \end{multline}
The sum is taken with respect to all partitions of the set $\bar t$  into
subsets $\bar t_{\so},\;\bar t_{\st}$  and the set $\bar x$ into subsets $\bar x_{\so},\;\bar x_{\st}$  with $\#\bar t_{\so}=
\#\bar x_{\so}=n$, $n=0,1,\dots,a$.

\item {\it The sum over partitions of $\bar s$ and $\bar y$.}
 \ba{S-GF}
  Z_{a,b}(\bar t;\bar x|\bar s;\bar y)=\sum (-1)^{n} f(\bar s_{\st},\bar t) f(\bar y_{\so},\bar x)
 f(\bar s_{\so},\bar s_{\st})f(\bar y_{\st},\bar y_{\so})\\
 \times K_n(\bar y_{\so}|\bar s_{\so}) K_{b-n}(\bar y_{\st}+c|\bar s_{\st})
  K_{a+n}(\bar s_{\so},\bar x|\bar y_{\so}+c,\bar t).
 \end{multline}

The sum is taken with respect to all partitions of the set $\bar s$  into
subsets $\bar s_{\so},\;\bar s_{\st}$  and the set $\bar y$ into subsets $\bar y_{\so},\;\bar y_{\st}$
 with $\#\bar s_{\so}=\#\bar y_{\so}=n$, $n=0,1,\dots,b$.

\end{itemize}

\subsection{Integral representations\label{S-Int-Part}}

Now we give several representations for $Z_{a,b}$ in terms
of multiple contour integrals of Cauchy type. The formulas in terms of sums
over partitions given above follow from the integral representations.

\begin{itemize}

\item $b$-fold integrals.
 \be{Int-Or-for}
 Z_{a,b}(\bar t;\bar x|\bar s;\bar y)=\frac1{(2\pi ic)^bb!}\oint\limits_{\bar w} K_b(\bar s-c|\bar z)
   K_b(\bar y|\bar z)K_{a+b}(\bar w|\bar t,\bar z+c)f(\bar z,\bar w)\mathcal{F}_b(\bar z)
   \,d\bar z,
   \ee
where
\be{CF}
\mathcal{F}_b(\bar z)=\prod_{j,k=1\atop{j\ne k}}^b f^{-1}(z_j,z_k),
\ee
and $d\bar z=dz_1,\dots,dz_b$. We have used a subscript $\bar w$ on  the integral symbol in order
to stress that the integration contour for every $z_j$ surrounds the set $\bar w=\{\bar s,\;\bar x\}$ in
the counterclockwise direction. We also assume that the integration contours do not contain any
other singularities of the integrand. Similar prescription will be kept for all other integral representations
considered below.

One more $b$-fold integral for $Z_{a,b}$ has the form
 \be{Int-Al-for}
Z_{a,b}(\bar t;\bar x|\bar s;\bar y)=\frac{(-1)^bf(\bar y,\bar x)f(\bar s,\bar t)}{(2\pi ic)^b b!}\oint\limits_{\bar{\tilde\eta}}
K_b(\bar z|\bar s)    K_b(\bar z|\bar y+c)K_{a+b}(\bar x,\bar z|\bar{\tilde\eta})
   f(\bar{\tilde\eta},\bar z)
   \mathcal{F}_b(\bar z)\,d\bar z.
   \ee
Here $\bar{\tilde\eta}=\{\bar y,\bar t-c\}$.

\item $a$-fold integrals.

 \be{Int-Or-for-twin}
 Z_{a,b}(\bar t;\bar x|\bar s;\bar y)=\frac{(-1)^a}{(2\pi ic)^aa!}\oint\limits_{\bar w}
 K_a(\bar z|\bar x+c)K_a(\bar z|\bar t)K_{a+b}(\bar y,\bar z-c|\bar w)
  f(\bar w,\bar z) \mathcal{F}_a(\bar z)\,d\bar z,
   \ee
where $d\bar z=dz_1,\dots,dz_a$. The integration contours  surround the set $\bar w$, like in \eqref{Int-Or-for}.

An analog of \eqref{Int-Al-for}  has the form
 \be{Int-Al-for-twin}
Z_{a,b}(\bar t;\bar x|\bar s;\bar y)=\frac{f(\bar y,\bar x)f(\bar s,\bar t)}{(2\pi ic)^aa!}\oint\limits_{\bar{\eta}}
 K_a(\bar t-c|\bar z)    K_a(\bar x|\bar z)K_{a+b}(\bar{\eta}-c|\bar s,\bar z)
    f(\bar z,\bar\eta)
   \mathcal{F}_a(\bar z)\,d\bar z.
   \ee
The integration contours  surround the set $\bar{\eta}=\{\bar y+c,\bar t\}$.

\end{itemize}

\subsection{Recursions for $Z_{a,b}$ \label{S-RecHC-3}}

The partition function defined by Fig.~\ref{F-ParFun-def} possesses several important properties. First, it is a symmetric
function with respect to any set of variables $\bar y$, $\bar x$,  $\bar s$, or  $\bar t$.
This property  follows from the Yang--Baxter equation \eqref{Y-B} (see e.g. \cite{Res86,Whe12}).

The second property is that $Z_{a,b}$ is a rational function decreasing at least as $1/z$ at $z\to\infty$, where $z$ is an arbitrary
argument of the partition function. This property is almost evident. Consider an arbitrary horizontal (or
vertical) line of the lattice.
Note that $a$- and $b$- type vertices behave as 1 as $z\to\infty$, while the $c$-type vertex behaves as $1/z$. Thus, it is enough to show that at least one $c$-vertex is on the line.
As there are different indices on the both sides of the line, moving along this line
we must meet a $c$-type vertex somewhere. The corresponding statistical weight decreases at
infinity.

The most important property of the partition function (or, what is the same, of the highest coefficient) is that the residues of $Z_{a,b}$ in its poles can be expressed in terms of $Z_{a-1,b}$ or $Z_{a,b-1}$. Since $Z_{a,b}$ is a rational function in all its variables, this property formally allows us to fix the partition function unambiguously, provided we know  $Z_{a,b}$ for small $a$ and $b$.  It is easy to see that for $a=0$ or $b=0$  $Z_{a,b}$ coincides with $K_n$:
\be{Z-small}
Z_{a,0}(\bar t;\bar x|\emptyset;\emptyset)=K_{a}(\bar x|\bar t),\qquad
Z_{0,b}(\emptyset;\emptyset|\bar s;\bar y)=K_b(\bar y|\bar s).
\ee
Thus, if we find the recursions of $Z_{a,b}$ in its poles, we will fix it completely.

Consider, for example, $Z_{a,b}$ as a function of $s_b$ with all other variables fixed.
Then it has simple poles at $s_b= y_m$, $m=1,\dots,b$ and  $s_b= t_\ell$, $\ell=1,\dots,a$. Due to the
symmetry of $Z_{a,b}$ over $\bar y$ and over $\bar t$ it is enough to
find the residues at $s_b=y_b$ and $s_b=t_a$.

\begin{prop}\label{Two-rec}
The residue of $Z_{a,b}$ at $s_b=y_b$ is expressed in terms of $Z_{a,b-1}$:
\be{Rec-Z-triv}
\Bigl.\Res Z_{a,b}(\bar t;\bar x|\bar s;\bar y)\Bigr|_{ s_b= y_b}=
-cf( y_b,\bar s_{\check b})f(\bar y_{\check b}, y_b)f( y_b,\bar x)
Z_{a,b-1}(\bar t;\bar x|\bar s_{\check b};\bar y_{\check b}).
\ee
Recall that $\bar s_{\check b}=\bar s\setminus s_{b}$,  $\bar y_{\check b}=\bar y\setminus y_{b}$.

The residue of $Z_{a,b}$ at $s_b=t_a$ is expressed in terms of $Z_{a-1,b}$:
\be{Rec-Z-nontriv}
\Bigl.\Res Z_{a,b}(\bar t;\bar x|\bar s;\bar y)\Bigr|_{ s_b= t_a}=
c f( \bar s_{\check b}, t_a)f( t_a, \bar t_{\check a})\sum_{p=1}^a g(x_p, t_a)f(\bar x_{\check p},x_p)
Z_{a-1,b}(\bar t_{\check a};\bar x_{\check p}|\{\bar s_{\check b},\; x_p\};\bar y_{b}).
\ee
Here $\bar x_{\check p}=\bar x\setminus x_{p}$.
\end{prop}

The proof is given in section \ref{S-prf-RecHC-3}.

{\sl Remark.} The highest coefficient $Z_{a,b}(\bar t;\bar x|\bar s;\bar y)$ also has poles at
$x_j=t_k$ and $x_j=y_k$. The residues in these poles satisfy recursions similar to
\eqref{Rec-Z-triv}, \eqref{Rec-Z-nontriv} (see e.g. \cite{Res86}). We do not present these
recursions explicitly, because we do not use them.

\section{Proofs of the sum formulas for $Z_{a,b}$\label{S-proof1}}

In this section we prove  representations \eqref{RHC-IHC} and \eqref{Al-RHC-IHC} for $Z_{a,b}$.

We begin with equation \eqref{RHC-IHC}.
It follows from the  properties
of $K_n$ that $Z_{a,b}$ defined by \eqref{RHC-IHC} is a symmetric function with respect to every
set of variables and goes to zero  as one of its arguments goes to infinity. The initial conditions \eqref{Z-small}
obviously are valid.
Hence, it is enough to prove that $Z_{a,b}$ defined by \eqref{RHC-IHC} and considered as a function of $s_b$ possesses the
following properties:

\begin{itemize}
\item it has poles only at $s_b=y_k$, $k=1,\dots,b$ and at $s_b=t_j$, $j=1,\dots,a$;

\item the residues in these poles satisfy the recursions established in the previous section.

\end{itemize}

First we find the poles of $Z_{a,b}$ defined by \eqref{RHC-IHC}.

Due to the product $f(\bar w_{\so},\bar w_{\st})$
every single term in \eqref{RHC-IHC} may have poles at $w_{j}= w_{k}$.
It is clear, however, that
these singularities cancel each other due to the sum over partitions. More precisely, the residue at each $w_{j}= w_{k}$ will vanish, due to opposite contribution of the terms $(w_j\in \bar w_{{\so}},w_k\in\bar w_{{\st}})$ and $(w_j\in\bar w_{{\st}},w_k\in\bar w_{{\so}})$.

Other poles of $Z_{a,b}$ should
coincide with the poles of the three $K_n$ terms entering \eqref{RHC-IHC}. If we set
$w_{\so}=\{\bar s_{\so},\;\bar x_{\st}\}$
and $\bar w_{\st}=\{\bar s_{\st},\;\bar x_{\so}\}$, then due to \eqref{Rav} we have
 \begin{equation}\label{Rav1}
 K_b(\bar s-c|\bar w_{\so})= (-1)^{b+k}  K_k(\bar s_{\st}-c|\bar x_{\st}).
 \end{equation}
This $K_k$ function has pole at $s_b=x_j+c$, if $s_b\in\bar s_{\st}$ and $x_j\in\bar x_{\st}$. However in this case
the product $f(\bar w_{\so},\bar w_{\st})$ contains the factor $f(x_j,s_b)$, which vanishes at $s_b=x_j+c$.
Hence,  $K_b(\bar s-c|\bar w_{\so})$ does not produce poles in the r.h.s. of \eqref{RHC-IHC}.

Thus, we conclude that the poles of $Z_{a,b}$ coincide with the poles of the two remaining $K_a(\bar w_{\st}|\bar t)$ and $K_b(\bar y|\bar w_{\so})$, which are just at the points $s_b=y_k$, $k=1,\dots,b$ and $s_b=t_j$, $j=1,\dots,a$. It remains to check
the recursions \eqref{Rec-Z-triv} and \eqref{Rec-Z-nontriv}.

We start with the first one: let  $ s_b\to y_b$. The pole occurs if and only if $ s_b\in \bar w_{\so}$. Let
$\bar w_{\so}=\{\bar w_{\rm i}, s_b\}$. Using \eqref{Rec-K} we find
 \be{Rec-K1}
\Bigl. \Res K_b(\bar y|\bar w_{\rm i},s_b)\Bigr|_{s_b=y_b}=
-c f(\bar y_{\check b}, y_b)f( y_b,\bar w_{\rm i}) K_{b-1}(\bar y_{\check b}|\bar w_{\rm i}).
\ee
Substituting this into \eqref{RHC-IHC} and using \eqref{K-K} we obtain
 \ba{RHC-t-pole-1}
 \Bigl.\Res  Z_{a,b}\Bigr|_{s_b=y_b}=(-1)^{b}c f(\bar y_{\check b}, y_b)
  {\sum}'  K_{b-1}(\bar s_{\check b}-c|\bar w_{\rm i})\\
  \times K_{b-1}(\bar y_{\check b}|w_{\rm i})K_a(\bar w_{\st}|\bar y)
  f(\bar w_{\rm i},\bar w_{\st})
  f( y_b,\bar w_{\rm i})
  f( y_b, \bar w_{\st})  ,
    \end{multline}
where $\sum'$ means that the sum is taken over partitions of the set $\bar w\setminus s_b$. It remains to observe that
\be{Ind-part}
f( y_b,\bar w_{\rm i})f( y_b, \bar w_{\st})=
f( y_b, \bar s_{\check b})f( y_b, \bar x),
\ee
independently on a specific partition. Hence, re-denoting $\bar w_{\rm i}$ by $\bar w_{\so}$ we obtain
 \ba{RHC-t-pole-2}
 \Bigl.\Res  Z_{a,b}\Bigr|_{s_b=y_b}=(-1)^{b} c
 f( y_b,\bar s_{\check b})f( \bar y_{\check b},y_b)f( y_b,\bar x)
 \\
\times \sum
 K_{b-1}(\bar s_{\check b}-c|\bar w_{\so}) K_{b-1}(\bar y_{\check b}|\bar w_{\so})
  K_a(\bar w_{\st}|\bar t)f(\bar w_{\so},\bar w_{\st}),
    \end{multline}
where now $\bar w=\{\bar s_{\check b},\;\bar x\}$. The sum over partitions evidently gives $Z_{a,b-1}$ with
$s_b$ and $ y_b$ omitted. We arrive at \eqref{Rec-Z-triv}.

Consider  now the residue of $Z_{a,b}$ at $ s_b\to t_a$. The pole occurs if and only if $ s_b\in \bar w_{\st}$. Let
$\bar w_{\st}=\{\bar w_{\rm ii},\;  s_b\}$. Due to \eqref{Rec-K} we have
 \be{Rec-K2}
\Bigl. \Res K_a(\bar w_{\rm ii},  s_b|\bar y)\Bigr|_{s_b=t_a}=c
f( t_a, \bar t_{\check a})f(\bar w_{\rm ii}, t_a) K_{a-1}(\bar w_{\rm ii}|\bar t_{\check a}).
\ee
Substituting this into \eqref{RHC-IHC} we find
 \ba{RHC-nt-pole-1}
 \Bigl. \Res Z_{a,b}\Bigr|_{s_b=t_a}=(-1)^b c f( t_a, \bar t_{\check a}) {\sum}'
 K_b(\bar s_{\check b}-c, t_a-c|\bar w_{\so})\\
  \times K_b(\bar y|\bar w_{\so})K_{a-1}(\bar w_{\rm ii}|\bar t_{\check a})
  f(\bar w_{\so},\bar w_{\rm ii})f(\bar w_{\rm ii}, t_a)f(\bar w_{\so}, t_a),
    \end{multline}
where $\sum'$ again means that the sum is taken over partitions of the set $\bar w\setminus s_b$.
Using
\be{Ind-part1}
f(\bar w_{\rm ii},t_a)f(\bar w_{\so},t_a)=
f(\bar s_{\check b},t_a)f(\bar x,t_a),
\ee
we obtain
 \ba{RHC-nt-pole-2}
 \Bigl. \Res Z_{a,b}\Bigr|_{s_b=t_a}=(-1)^b c f( t_a, \bar t_{\check a})
f( \bar s_{\check b}, t_a) f(\bar x, t_a)
{\sum}' K_b(\bar s_{\check b}-c, t_a-c|\bar w_{\so})
   \\
  \times K_b(\bar y|\bar w_{\so})K_{a-1}(\bar w_{\rm ii}|\bar t_{\check a})
  f(\bar w_{\so},\bar w_{\rm ii}).
    \end{multline}
Observe that
 \be{Combi}
 f(\bar x, t_a)K_b(\bar s_{\check b}-c, t_a-c|\bar w_{\so})\to 0,\quad
 \mbox{at}\quad  t_a\to\infty,
 \ee
and this combination as a function of $t_a$ has poles only at $ t_a=x_p$, $p=1,\dots,a$. Hence, developing
it over these poles we have
 \be{Combi-1}
 f(\bar x, t_a)K_b(\bar s_{\check b}, t_a-c|\bar w_{\so})
 = \sum_{p=1}^ag(x_p, t_a) f(\bar x_{\check p},x_p) K_b(\bar s_{\check b}-c, x_p-c|\bar w_{\so}).
 \ee
Substituting this into \eqref{RHC-nt-pole-2}  and re-denoting $\bar w_{\rm ii}$ by $\bar w_{\st}$ we arrive at
 \ba{RHC-nt-pole-3}
 \Bigl. \Res Z_{a,b}\Bigr|_{s_b=t_a}=(-1)^b c
 f( t_a, \bar t_{\check a})f( \bar s_{\check b}, t_a)
\sum_{p=1}^ag(x_p, t_a) f(\bar x_{\check p},x_p)
   \\
  \times\sum  K_b(\bar s_{\check b}-c, x_p-c|\bar w_{\so})
  K_b(\bar y|\bar w_{\so})
  K_{a-1}(\bar w_{\st}|\bar t_{\check a})
  f(\bar w_{\so},\bar w_{\st}) ,
    \end{multline}
where now $\bar w=\{\bar s_{\check b},\;\bar x\}$. We see that the sum over the partitions gives exactly the highest
coefficient
$Z_{a-1,b}(\bar t_{\check a};\bar x_{\check p}|\{\bar s_{\check b}; x_p\},\bar y)$ and we reproduce
\eqref{Rec-Z-nontriv}. This ends the proof of relation \eqref{RHC-IHC}.

\medskip

Consider now $ Z_{a,b}$ defined by equation \eqref{Al-RHC-IHC}. The recursion in the pole at $s_b=y_b$ for this representation can be checked in a manner similar to the one described above. The proof of the recursion in the pole at $s_b=t_a$ is slightly
different. This pole is in the product $f(\bar s,\bar t)$ in the
formula \eqref{Al-RHC-IHC}. We have
 \ba{twin-nt1}
  \Res Z_{a,b}\Bigr|_{s_b=t_a}= c f(\bar s_{\check b},t_a)f(t_a,\bar t_{\check a})
 f(\bar s_{\check b},\bar t_{\check a})f(\bar y,\bar x)\\
  \times(-1)^a\sum
 K_a(\bar t-c|\bar\eta_{\so})K_a(\bar x|\bar\eta_{\so})
  K_b(\bar\eta_{\st}-c|\bar s_{\check b}, t_a)f(\bar\eta_{\so},\bar\eta_{\st}).
    \end{multline}

Consider the second line of \eqref{twin-nt1} as a rational function of $t_a$. This rational function evidently vanishes as $t_a\to\infty$. Let us find the poles of this function. Suppose that
$t_a\in\bar\eta_{\st}$. Then due to \eqref{K-K} $t_a$ disappears from the arguments of $K_b(\bar\eta_{\st}-c|\bar s_{\check b}, t_a)$,
but it remains in $K_a(\bar t-c|\bar\eta_{\so})$. There might be poles if $t_a-c$ coincides with some
element belonging to $\eta_{\so}$,
but they are compensated by the zeros of the product $f(\bar\eta_{\so},\bar\eta_{\st})$. Hence, the
rational function has no poles in this case.

Let now $t_a\in\bar\eta_{\so}$. Then $t_a$ disappears from $K_a(\bar t-c|\bar\eta_{\so})$,
but it appears in $K_a(\bar x|\bar\eta_{\so})$, where we obtain the poles at $t_a=x_p$, $p=1,\dots,a$.
Introducing $\bar\eta_{\rm i}=\bar\eta_{\so}\setminus t_a$ and using \eqref{K-sum-x}, \eqref{K-K}, we develop
$K_a(\bar x|\bar\eta_{\so})$ over the poles at $t_a=x_p$ to rewrite \eqref{twin-nt1} as
 \ba{twin-nt2}
 \Res  Z_{a,b}\Bigr|_{s_b=t_a}=(-1)^{a-1}c f(\bar s_{\check b},t_a)f(t_a,\bar t_{\check a})
  f(\bar s_{\check b},\bar t_{\check a})f(\bar y,\bar x){\sum}'\sum_{p=1}^a
 K_{a-1}(\bar t_{\check a}-c|\bar\eta_{\rm i})\\
  \times  g(x_p,t_a)f(\bar x_{\check p},x_p)
  f(x_p,\bar\eta_{\rm i}) K_{a-1}(\bar x_{\check p}|\bar\eta_{\rm i})
  K_b(\bar\eta_{\st}-c|\bar s_{\check b},x_p)f(\bar\eta_{\rm i},\bar\eta_{\st})
    f(x_p,\bar\eta_{\st}),
    \end{multline}
where $\sum'$ means that the sum is taken over the partitions of the set $\bar\eta\setminus t_a$ into the subsets $\bar\eta_{\rm i}$ and $\bar\eta_{\st}$. Using
 \be{prod f1}
   f(x_p,\bar\eta_{\rm i})f(x_p,\bar\eta_{\st})=
f(x_p,\bar t_{\check a})f(x_p,\bar y+c)=f(x_p,\bar t_{\check a})f^{-1}(\bar y,x_p),
  \ee
and re-denoting $\bar\eta_{\rm i}=\bar\eta_{\so}$ we re-write \eqref{twin-nt2} in the form
 \ba{twin-nt3}
 \Res Z_{a,b}\Bigr|_{s_b=t_a}=c f(\bar s_{\check b},t_a)f(t_a,\bar t_{\check a})
  \sum_{p=1}^a g(x_p,t_a)f(\bar x_{\check p},x_p)\\
  \times f(\bar s_{\check b},\bar t_{\check a})f(x_p,\bar t_{\check a})
f(\bar y,x_{\check p})
  (-1)^{a-1} {\sum}
 K_{a-1}(\bar t_{\check a}-c|\bar\eta_{\so})     K_{a-1}(\bar x_{\check p}|\bar\eta_{\so})
      K_b(\bar\eta_{\st}-c|\bar s_{\check b},x_p) f(\bar\eta_{\so},\bar\eta_{\st}) .
    \end{multline}
Looking at the expression in the second line of \eqref{twin-nt3} one can easily recognize the highest
coefficient
$Z_{a-1,b}(\bar t_{\check a};\bar x_{\check p}|\{\bar s_{\check b},x_p\};\bar y)$ defined by \eqref{Al-RHC-IHC}.

One can also prove that the representations \eqref{GF}, \eqref{S-GF} satisfy the recursions \eqref{Rec-Z-triv} and
\eqref{Rec-Z-nontriv}. We will use, however, another method based on the multiple integral representations  for $Z_{a,b}$.

\section{Proofs of the integral representations for $Z_{a,b}$\label{S-proof2}}

All the integral representations listed in subsection~\ref{S-Int-Part} can be proved in a
similar manner. Namely, they can be reduced to the sums over partitions listed in Section~\ref{S-Rep-Part}.

Consider  for example \eqref{Int-Or-for-twin}. The only poles of
the integrand within the integration contours are the points $z_j=w_k$. Evaluating the integral
by the residues in these poles we obtain
 \be{Int-Or-for-twin-res}
 Z_{a,b}=\sum K_a(\bar w_{\st}-c|\bar x)K_a(\bar w_{\st}|\bar t)K_{a+b}(\bar y,\bar w_{\st}-c|\bar w)
  f(\bar w_{\so},\bar w_{\st}),
   \ee
where the sum is taken over partitions of $\bar w$ into subsets $\bar w_{\so}$ and $\bar w_{\st}$ with
$\#\bar w_{\so}=b$ and $\#\bar w_{\st}=a$. Due to
\eqref{K-K} we have
\be{red-Kab}
K_{a+b}(\bar y,\bar w_{\st}-c|\bar w)=(-1)^aK_{b}(\bar y|\bar w_{\so}),
\ee
and we immediately arrive at \eqref{RHC-IHC-twin}.

Dealing with contour integrals of  rational functions we always have a possibility to calculate them
by the residues in the poles outside the original integration contour. Consider the integrand in
\eqref{Int-Or-for-twin} as a function of some $z_j$. It behaves as $1/z_j^3$ at $z_j\to\infty$, hence, the
residue at infinity vanishes. The  poles outside the original integration contour are in the points $z_j=t_k$ and $z_j=x_k+c$
(the poles at $z_j=w_k+c$ are compensated by the zeros of the product $f(\bar w, \bar z)$). Due to the factor
$\mathcal{F}_a(\bar z)$ the integrand also has poles at $z_j=z_k\pm c$  for $k\ne j$. These last poles do not
contribute to the final result (see Appendix~\ref{A-UP}), thus, we can move the original
contour surrounding $\bar z=\bar w$ to the points $\bar z=\bar t$ and $\bar z=\bar x+c$
 \be{Int-Or-for-twin1}
 Z_{a,b}=\frac{1}{(2\pi ic)^aa!}\oint\limits_{\bar\xi}
 K_a(\bar z|\bar x+c)K_a(\bar z|\bar t)K_{a+b}(\bar y,\bar z-c|\bar w)
  f(\bar w,\bar z) \mathcal{F}_a(\bar z)\,d\bar z,
   \ee
where we have combined the sets $\bar t$ and $\bar x+c$ into one set $\bar\xi$.
Now we can compute this integral by the residues using \eqref{Rec-K}. However, it is more
convenient first to transform slightly the integrand. Namely, applying \eqref{Red-K}
for all three $K_n$ in \eqref{Int-Or-for-twin1} we obtain
 \be{Int-Or-for-twin2}
 Z_{a,b}=\frac{(-1)^{a+b}}{(2\pi ic)^aa!}\oint\limits_{\bar\xi}
 K_a(\bar t-c|\bar z)K_a(\bar x|\bar z)K_{a+b}(\bar w|\bar y+c,\bar z)
  f(\bar y,\bar w)f(\bar z,\bar\xi) \mathcal{F}_a(\bar z)\,d\bar z.
   \ee
Now all the poles are explicitly combined in the factor $f(\bar z,\bar\xi)$. Hence, the
result of the integration gives the sum over partitions of $\bar\xi$ into $\bar\xi_{\so}$
and $\bar\xi_{\st}$ with $\#\bar\xi_{\so}=\#\bar\xi_{\st}=a$:
 \be{Int-Or-for-twin3}
 Z_{a,b}=(-1)^{a+b}\sum
 K_a(\bar t-c|\bar\xi_{\so})K_a(\bar x|\bar\xi_{\so})K_{a+b}(\bar w|\bar y+c,\bar\xi_{\so})
  f(\bar y,\bar w)f(\bar\xi_{\so},\bar\xi_{\st}).
 \ee
It remains to set
 \be{xi-xt}
 \begin{array}{l}
 \bar\xi_{\so}= \{\bar t_{\so},\;\bar x_{\st}+c\},\\
 \bar\xi_{\st}= \{\bar t_{\st},\;\bar x_{\so}+c\},
 \end{array}
 \qquad \#\bar t_{\so}=\#\bar x_{\so}=n,\quad n=0,\dots,a,
 \ee
and after simple algebra we arrive at \eqref{GF}. Thus, the two sum formulas \eqref{RHC-IHC-twin} and \eqref{GF} are
different representations of the same integral \eqref{Int-Or-for-twin}. Since the equation \eqref{RHC-IHC-twin}
was already proved, we automatically obtain the proof of the equation \eqref{GF}.

Similarly one can check that direct evaluation of the integrals by the residues within the original contours in the representations \eqref{Int-Or-for}, \eqref{Int-Al-for}, and \eqref{Int-Al-for-twin}
give respectively the sum formulas \eqref{RHC-IHC}, \eqref{Al-RHC-IHC-twin}, and \eqref{Al-RHC-IHC}.
The sum formula \eqref{S-GF} follows, for example, from \eqref{Int-Or-for} after moving the contours in this representation.

\section{Proof of the recursion formulas for $Z_{a,b}$ \label{S-prf-RecHC-3}}
The recursion \eqref{Rec-Z-triv} was pointed out already in \cite{Res86} (see also \cite{Whe12}).
Therefore we give only a sketch of the proof for this relation, while detailing the proof for the recursion relation \eqref{Rec-Z-nontriv}.

We start wih relation  \eqref{Rec-Z-triv}. Due to the symmetry
of the partition function with respect to each set of variables, we can assume that the parameter $ y_b$
corresponds to the extreme left vertical line, while the parameter $s_b$
corresponds to the extreme lower horizontal line (see Fig.~\ref{F-ParFun-def}).
Then the pole at $ s_b= y_b$ occurs if and only if the
extreme South-West vertex is of $c$-type. It is easy to see that as soon as the $c$-type of the extreme South-West vertex is fixed, all the vertices along the left and lower boundaries can be restored unambiguously. Namely, the vertices corresponding to the variables $(s_b,\bar y_{\check b})$, $(\bar s_{\check b},y_b)$, and $(\bar x,y_b)$ are of $a$-type, while the vertices corresponding to the variables $(s_b,\bar t)$ are of $b$-type. The product of the corresponding statistical weights gives us the prefactor in \eqref{Rec-Z-triv}. The remaining sub-lattice is $Z_{a,b-1}$, in which $s_b$ and $y_b$ are excluded. Thus, we arrive at \eqref{Rec-Z-triv}.

The recursion \eqref{Rec-Z-nontriv} is slightly more sophisticated.
Let, as before, $ s_b$ correspond to the lower horizontal
line, while $ t_a$ corresponds to the right vertical line. Then the pole at $ s_b= t_a$
occurs if and only if the extreme South-East vertex is of $c$-type. Then we can restore all the vertices along the lower
horizontal line and a part of vertices along the right vertical line. Namely, the vertices corresponding to the variables
$(s_b,\bar y_{\check b})$ are of $b$-type, while the vertices corresponding to the variables $(\bar s_{\check b},t_a)$ and
$(s_b,\bar t_{\check a})$ are of $a$-type. The product of the corresponding statistical weights is
$g( s_b, t_a)f( s_b, \bar t_{\check a})f( \bar s_{\check b}, t_a)$. Thus, we obtain
\be{Non-square}
\Bigl. \Res Z_{a,b}\Bigr|_{ s_b= t_a}=cf( t_a, \bar t_{\check a})f( \bar s_{\check b}, t_a)
\cdot \widetilde Z_{a,b},
\ee
where $\widetilde Z_{a,b}$ is given as a new partition function shown on Fig.~\ref{F-ParFun-Sub}.

\begin{figure}[h!]
\noindent
\hspace{3cm}\parbox[b][34mm][t]{40mm}{$\dis \tilde Z_{a,b}(\bar t;\bar x|\bar s_{\check b};\bar y)\quad=$}
\parbox[b][6cm][t]{40mm}{\begin{picture}(200,200)
\put(40,-20){%
\begin{picture}(200,200)
\multiput(20,30)(30,0){5}{\line(0,1){140}}
\multiput(10,40)(0,30){2}{\line(1,0){140}}
\multiput(10,100)(0,30){3}{\line(1,0){170}}
\multiput(16,36)(0,30){5}{$\bullet$}
\put(170,90){\line(0,1){80}}
\multiput(46,36)(0,30){5}{$\bullet$}
\multiput(76,36)(0,30){5}{$\bullet$}
%
%
\multiput(4,38)(0,30){5}{$\scriptstyle 2$}
%
\multiput(183,98)(0,30){3}{$\scriptstyle 1$}
\multiput(23,172)(30,0){3}{$\scriptstyle 2$}
\multiput(113,172)(30,0){3}{$\scriptstyle 1$}
\put(110,180){$\overbrace{\hphantom{\hspace{1.5cm}}}^{\dis \bar t}$}
\put(20,180){$\overbrace{\hphantom{\hspace{1.5cm}}}^{\dis \bar y}$}
\put(187,50){$\left.\rule{0mm}{5mm}\right\}{ \bar s_{\check b}}$}
\put(187,125){$\left.\rule{0mm}{9mm}\right\}{\bar x}$}
%
%
\put(173,83){$\scriptstyle 3$}       %
\multiput(155,42)(0,30){2}{$\scriptstyle 3$}
\multiput(23,23)(30,0){3}{$\scriptstyle 3$}
\multiput(113,23)(30,0){2}{$\scriptstyle 2$}
%
\end{picture}
}
\end{picture}}
\caption{\label{F-ParFun-Sub} The partition function $\tilde Z_{a,b}(\bar t;\bar x|\bar s_{\check b};\bar y)$.
The sublattice is obtained from the original lattice by removing the vertices $(s_b,\bar y)$, $(s_b,\bar t)$
and $(\bar s_{\check b}, t_a)$.}
\end{figure}

Consider the remaining partition function as a function of $ t_a$: $\widetilde Z_{a,b}=\widetilde Z_{a,b}( t_a)$.
This variable corresponds to the extreme right vertical line (the shorter one) of the lattice.
The indices at the ends of this line are different, hence, a $c$-type vertex should be somewhere
on this  line. Then
\be{tZ-poles2}
\widetilde Z_{a,b}( t_a)=\sum_{p=1}^a \Gamma_p \;g(x_p, t_a),
\ee
where $\Gamma_p$ do not depend on $ t_a$. Due to the symmetry of $\widetilde Z_{a,b}$ over $\bar x$
it is enough to find only $\Gamma_a$. All other coefficients $\Gamma_p$ can be obtained from $\Gamma_a$ via the
replacement $x_a\leftrightarrow x_p$. The contribution from the term $p=a$ in \eqref{tZ-poles2} occurs if and only if the lowest
vertex on the short line is of $c$-type. Then all the remaining vertices on the short line
are of $a$-type, while the $t_a$-independent  coefficient is equal to $Z_{a-1,b}$, where $ s_b$ is replaced by $x_a$.
Thus, we find
\be{Gamma-a}
\Gamma_a= Z_{a-1,b}(\bar t_{\check a},\bar x_{\check a}|\{\bar s_{\check b},\; x_a\},\bar y)
f(\bar x_{\check a},x_a),
\ee
and hence,
\be{Gamma-p}
\Gamma_p= Z_{a-1,b}(\bar t_{\check a},\bar x_{\check p}|\{\bar s_{\check b},\; x_p\},\bar y)
f(\bar x_{\check p},x_p).
\ee
Substituting this into \eqref{tZ-poles2} and using \eqref{Non-square} we prove the recursion \eqref{Rec-Z-nontriv}.

\medskip
Starting from $Z_{0,b}$ and $Z_{a,0}$ and using the recursions \eqref{Rec-Z-triv}, \eqref{Rec-Z-nontriv} we can
construct iteratively $Z_{1,b}$ for $b=1,2\dots$, then $Z_{2,b}$ for $b=2,3\dots$ and so on.

\section*{Conclusion}

We have presented several different formulas for the highest coefficient $Z_{a,b}$. We hope that at least
some of them will be useful for further applications.

There exists at least one more representation for $Z_{a,b}$ that we did not mention in this paper.
It is analogous to the equation \eqref{Sym1} for $K_n(\bar x|\bar y)$. Such type of representations
naturally appeared in the series of papers devoted to the universal description of the nested Bethe vectors
(see \cite{KhP} and references therein). In this approach the Bethe vectors are
expressed through the modes of the generating series  in the `current' realization
of a quantum affine algebra or Yangian double. The rational
function $K_n(\bar x|\bar y)$ serves as a kernel of the integral transform which
relates Bethe vectors with product of currents of the Yangian double (in case of the models with
$SU(2)$-symmetry). In \cite{BelPR10}, analogous kernel for the integral transform in the `current' approach to
the universal Bethe vectors was constructed for the model with
$U_q(\mathfrak{gl}_3)$-symmetry.
Since such kernel can be naturally associated with highest coefficient,
the  rational limit of the kernels found in \cite{BelPR10} yields one more representation of
$Z_{a,b}$. The corresponding proof  will be published elsewhere.

Turning back to the representations considered in this paper we should note that all of them are given in terms of sums over partitions or in terms of multiple integrals. Of course, it would be better to have a representation for $Z_{a,b}$ in terms of a
single determinant, as it has been done for $K_n$ in the $\frak{gl}_2$ case. However the structure of our formulas leads us
to conjecture that such a single determinant representation hardly exists.

If our conjecture is correct, then a determinant formula for the scalar product of a generic
Bethe vector with the transfer-matrix eigenvectors should not exist. Indeed, since $Z_{a,b}$
is a particular case of such scalar product, the existence of a (single) determinant formula for $Z_{a,b}$ is a prerequisite for the existence of a determinant formula for the scalar product. This negative result, however,
does not mean that there is no determinant representations for some cases of  scalar
products involving generic Bethe vectors. Some of such particular cases were considered in \cite{Whe12}.
In our forthcoming publication \cite{BelPRS12} we will present one more determinant formula, which allows to
calculate some form factors of local operators of the quantum XXX $SU(3)$-invariant Heisenberg chain.

\section*{Acknowledgements}
Work of S.P. was supported in part by RFBR grant 11-01-00980-a, grant
of Scientific Foundation of NRU HSE ╣ 12-09-0064 and grant of
FASI RF 14.740.11.0347.
N.A.S. was  supported by the Program of RAS Basic Problems of the Nonlinear Dynamics,
RFBR-11-01-00440, RFBR-11-01-12037-ofi-m, SS-4612.2012.1.

\appendix

\section{Properties of  $K_n$\label{A-IHC}}

 $K_n$ is  symmetric function of $x_1,\dots,x_n$ and symmetric function of $y_1,\dots,y_n$.
It behaves as $1/x_n$ (resp. $1/y_n$) as $x_n\to\infty$ (resp. $y_n\to\infty$) at other variables fixed.
It has simple poles at $x_j=y_k$. The residues in these poles can be expressed in terms of $K_{n-1}$:
 \be{Rec-K}
 \begin{array}{l}
\Bigl. \Res K_n(\bar x|\bar y)\Bigr|_{x_n= y_n}= -c
 f(y_n,\bar y_{\check n})f(\bar x_{\check n},y_n)\cdot K_{n-1}(\bar x_{\check n}|\bar y_{\check n}),\num
\Bigl.\Res K_n(\bar x|\bar y)\Bigr|_{y_n=x_n}= c
f(x_n,\bar y_{\check n})f(\bar x_{\check n},x_n)\cdot K_{n-1}(\bar x_{\check n}|\bar y_{\check n}).
\end{array}
\ee
Using \eqref{Rec-K} we can develop $K_n(\bar x|\bar y)$ with respect to the poles at $y_n=x_p$, $p=1,\dots,n$:
\be{K-sum-x}
K_n(\bar x|\bar y)= \sum_{p=1}^n g(x_p,y_n)
f(x_p,\bar y_{\check n})f(\bar x_{\check p},x_p) K_{n-1}(\bar x_{\check p}|\bar y_{\check n}).
\ee
The equation \eqref{K-sum-x} expresses $K_n$ in terms of $K_{n-1}$. Continuing this process
we arrive at the following representation
\be{Sym1}
K_n(\bar x|\bar y)=\mathop{\rm Sym}_{\bar x} \prod_{j=1}^n g(x_j,y_j)\prod_{j>k}^{n}
f(x_j,y_k)f(x_k,x_j),
\ee
where $\mathop{\rm Sym}_{\bar x}$ means the symmetrization over $\bar x$.

One can also easily check that  $K_n$ possesses the following properties:
 \be{K-K}
K_{n+1}(\bar x, z-c|\bar y, z)=K_{n+1}(\bar x, z|\bar y, z+c)= - K_{n}(\bar x|\bar y),
\ee
and
 \be{Red-K}
K_{n}(\bar x-c|\bar y)=K_{n}(\bar x|\bar y+c)= (-1)^n f^{-1}(\bar y,\bar x) K_{n}(\bar y|\bar x).
\ee

\begin{prop}\label{PropA} Let $x_2=x_1-c$. Then $K_n(\bar x|\bar y)$ as a function of $x_1$ is holomorphic in
the points $\bar y$.
\end{prop}

{\sl Proof.} If $x_2=x_1-c$, then the functions $t(x_1,y_k)$ and $t(x_2,y_k)$ have
poles at $x_1=y_k$, $k=1,\dots,n$ (see \eqref{K-def}). However the prefactor $h(\bar x,\bar y)$ contains
the product $h(x_2,\bar y)=g^{-1}(x_1,\bar y)$, that compensates these poles.

Due to the symmetry of $K_n$ over $\bar x$ the same property holds if $x_k=x_j-c$ for arbitrary $j$ and $k$.
Then $K_n(\bar x|\bar y)$ is a holomorphic function of $x_j$ in $\bar y$.

\section{Spurious poles\label{A-UP}}

Consider the integral representation \eqref{Int-Or-for-twin}. The integrand is a symmetric
function of the integration variables $\bar z$. Hence,  replacing   $K_a(\bar z|\bar t)$ via \eqref{Sym1}
we obtain
 \be{A-Int1}
 Z_{a,b}=\frac{(-1)^a}{(2\pi ic)^a}\oint\limits_{\bar w}\prod_{j=1}^ag(z_j,t_j)
 \prod_{j>k}^a \bigl(f(z_j,t_k)f^{-1}(z_j,z_k)\bigr)\cdot
 K_a(\bar z|\bar x+c)K_{a+b}(\bar y,\bar z-c|\bar w)
  f(\bar w,\bar z) \,d\bar z.
   \ee
The original integration contours surround the points $\bar w= \{\bar s,\;\bar x\}$. Our goal is to
move these contours to the points $\bar\xi= \{\bar t,\;\bar x+c\}$. We do it successively starting
from the contour for $z_1$.

Consider the integrand in \eqref{A-Int1} as a function of $z_1$. It  has poles at $z_1=t_1, x_1+c,\dots, x_a+c$
and at $z_1=z_2+c,\dots,z_a+c$ (recall that the poles at $z_1=w_m+c$ are compensated by the zeros of
the product $f(\bar w,\bar z)$). Consider the residue at $z_1=z_p+c$. Then
 \be{can-sing1}
 \Bigl.f(\bar w, z_1)f(\bar w, z_p)\Bigr|_{z_1=z_p+c}= f^{-1}( z_p,\bar w)f(\bar w, z_p)=\prod_{k=1}^{a+b}
 \frac{w_k-z_p+c}{w_k-z_p-c}.
 \ee
We see that the integrand as a function of $z_p$ becomes holomorphic at $z_p=w_k$, $k=1,\dots,a+b$. Hence,
the integral over $z_p$ vanishes, because the integration contour for $z_p$ still surrounds the points
$\bar w$. We conclude that the integration contour for $z_1$ can be moved to the points $t_1$ and $\bar x+c$
without any additional contribution from the poles at $z_1=z_p+c$ for $p=2,\dots,a$. We arrive at\footnote[1]{%
The obtained contour around  the points $t_1$ and $\bar x+c$ is oriented in the clockwise direction. Changing
the orientation of this contour we obtain in \eqref{A-Int2} the sign $(-1)^{a-1}$ instead of $(-1)^{a}$ in
\eqref{A-Int1}.}

 \begin{multline}\label{A-Int2}
 Z_{a,b}=\frac{(-1)^{a-1}}{(2\pi ic)^a}\oint\limits_{t_1,\bar x+c}\,dz_1\oint\limits_{\bar w}
 \,dz_2\dots \,dz_a
 \prod_{j=1}^ag(z_j,t_j)
 \prod_{j>k}^a \bigl(f(z_j,t_k)f^{-1}(z_j,z_k)\bigr)\\
 \times K_a(\bar z|\bar x+c)K_{a+b}(\bar y,\bar z-c|\bar w)
  f(\bar w,\bar z).
   \end{multline}

Now we move the integration contour for $z_2$ to the points $t_1, t_2$ and $\bar x+c$. Similarly to the
case considered above we do not obtain contributions from the poles at $z_2=z_p+c$ for $p=3,\dots,a$. One
additional pole arises at $z_2=z_1-c$. However, it is easy to check that taking the residue in this pole we make the integral over $z_1$ vanishing. Indeed
\be{canc-2}
\Bigl.g(z_1,t_1)f(z_2,t_1)\Bigr|_{z_2=z_1-c}=g(z_1,t_1)f^{-1}(t_1,z_1)=\frac{c}{z_1-t_1-c},
\ee
thus, the pole at $z_1=t_1$ disappears. On the other hand due to Proposition~\ref{PropA},
$K_a(\bar z|\bar x+c)$ at $z_2=z_1-c$ becomes  a holomorphic function of $z_1$ in the points $\bar x+c$.
Thus, all the integrand as a function of $z_1$ has no poles within the integration contour and therefore
the integral vanishes. We conclude that the contour for $z_2$ can be moved to the points $t_1, t_2$ and $\bar x+c$
without any additional contribution.

The process obviously can be continued. It is clear that setting $z_k=z_\ell+c$ for $\ell=k+1,\dots, b$ or
$z_k=z_\ell-c$ for $\ell=1,\dots, k-1$ we always obtain a  function of $z_\ell$, which is holomorphic in the domain
of the integration. Hence, the integral over $z_\ell$ vanishes.
Thus,  we arrive at
 \begin{multline}\label{A-Int3}
 Z_{a,b}=\frac{1}{(2\pi ic)^a}\oint\limits_{t_1,\bar x+c}\,dz_1\oint\limits_{t_1,t_2,\bar x+c}\,dz_2
 \dots\oint\limits_{\bar t,\bar x+c}\,dz_a
 \prod_{j=1}^ag(z_j,t_j)
 \prod_{j>k}^a \bigl(f(z_j,t_k)f^{-1}(z_j,z_k)\bigr)\\
 \times K_a(\bar z|\bar x+c)K_{a+b}(\bar y,\bar z-c|\bar w)
  f(\bar w,\bar z).
   \end{multline}
Obviously the integration contours for every $z_k$ can be extended to the contour surrounding
all the points $\bar t$, since the integrand as a function of $z_k$ is holomorphic in $t_{k+1},\dots,t_a$.
After that we symmetrize the integrand over all integration variables using
\eqref{Sym1}  and we finally obtain \eqref{Int-Or-for-twin1}.

\end{document}